**Enhanced interlayer neutral excitons and trions in trilayer van der Waals heterostructures**

Chanyeol Choi[1,2,9,†], Jiahui Huang[1,2], Hung-Chieh Cheng[3,4], Hyunseok Kim[2], Abhinav Kumar Vinod[1,2], Sang-Hoon Bae[3,4,10], V. Ongun Özçelik[5], Roberto Grassi[6], Jongjae Chae[3], Shu-Wei Huang[1,2,11], Xiangfeng Duan[4,7], Kristen Kaasbjerg[8], Tony Low[6], and Chee Wei Wong[1,2,†]

[1] Fang Lu Mesoscopic Optics and Quantum Electronics Laboratory, University of California, Los Angeles, CA 90095, United States

[2] Department of Electrical Engineering, University of California, Los Angeles, CA 90095, United States

[3] Department of Materials Science and Engineering, University of California, Los Angeles, CA 90095, United States

[4] California Nanosystems Institute, University of California, Los Angeles, CA 90095, United States

[5] Andlinger Center for Energy and the Environment, Princeton University, Princeton, New Jersey 08544, United States

[6] Department of Electrical and Computer Engineering, University of Minnesota, Minneapolis, MN 55455, United States

[7] Department of Chemistry and Biochemistry, University of California, Los Angeles, CA 90095, United States

[8] Center for Nanostructured Graphene (CNG), Department of Micro- and Nanotechnology (DTU Nanotech), Technical University of Denmark, DK-2800 Kgs. Lyngby, Denmark

[9] Present address: Department of Electrical Engineering and Computer Science, Massachusetts Institute of Technology, Cambridge, MA 02139, United States

[10] Present address: Department of Mechanical Engineering, Massachusetts Institute of Technology, Cambridge, MA 02139, United States

[11] Present address: Department of Electrical, Computer, and Energy Engineering, University of Colorado Boulder, Boulder, CO 80309, United States

[†] Corresponding authors: cowellchoi@gmail.com ; cheewei.wong@ucla.edu

**Vertically stacked van der Waals heterostructures constitute a promising platform for providing tailored band alignment with enhanced excitonic systems. Here we report the first observations of neutral and charged interlayer excitons in trilayer $WSe_2$-$MoSe_2$-$WSe_2$ van der Waals heterostructures and their dynamics. The addition of a $WSe_2$ layer in the trilayer leads to significantly higher photoluminescence quantum yields and tunable spectral resonance compared to its bilayer heterostructures at cryogenic temperatures. The observed enhancement in the photoluminescence quantum yield is due to significantly larger electron-hole overlap and higher light absorbance in the trilayer heterostructure, supported via first-principle pseudopotential calculations based on spin-polarized density functional theory. We further uncover the temperature- and power-dependence, as well as time-resolved photoluminescence of the trilayer heterostructure interlayer neutral excitons and trions. Our study elucidates the prospects of manipulating light emission from interlayer excitons and designing atomic heterostructures from first-principles for optoelectronics.**

In two-dimensional (2D) materials, Coulomb-induced electronic states of excitons have been examined as a platform to understand many-body carrier-carrier interactions. These excitonic interactions dominate in layered materials due to quantum confinement and reduced dielectric screening. 2D atomic crystals of transition metal dichalcogenides (TMDs) have provided new opportunities in the studies of single-exciton single-photon interactions, spin-orbit coupling, ultrafast dynamics, and nanoelectronic devices [1-22]. Layer-by-layer stacking of TMDs-based van der Waals (vdWs) heterostructures has recently captured the attention of the scientific community where the tailored band alignment with diverse 2D materials can be achieved via advanced 2D growth and transfer techniques [23-28]. This has allowed the extension of indirect excitons with spatially separated electrons and holes from, for example, coupled quantum wells in GaAs/AlGaAs [29,30] to TMD vdW heterostructure bilayers, where they are referred to as interlayer excitons. These interlayer excitons exhibit rich physics in TMDs-based vdWs heterostructures due to the novel atomic granularity control through layer-by-layer stacking and the chiral properties of quantum electronic states [8,31,32]. Recently several studies observed that heterostructure interlayer excitons feature long lifetimes, spin-valley polarization by circularly-polarized pumping, near-unity valley polarization via large magnetic splitting, and charge transfer at the heterogeneous interfaces [32-45].

Here we demonstrate TMDs-based vdWs heterostructures composed of a three-layer $WSe_2$-$MoSe_2$-$WSe_2$ stack to achieve unique band alignment that promotes efficient interlayer radiative recombination. We first design the heterostructure through pseudopotential calculations based on spin-polarized density functional theory, with the exchange-correlation potential estimated via the Perdew-Burke-Ernzerhof functional. Optimizing the designed heterostructure band structure through conjugate gradients, we include the effects of spin-orbit coupling through non-collinear computations. We correlate the obtained real-space pseudo-wavefunctions with lattice points in reciprocal space to understand the carrier localization, the type-II optical transition oscillator strength, and potential asymmetries in the electron and hole band profiles towards neutral interlayer excitons and three-body trion complexes. Enabled by the first-principle computations, we fabricate the trilayer heterostructure stack through hexagonal boron nitride assisted transfer. Our dry transfer process allows a relatively undoped heterostructure and atomic smoothness at the interfaces. We perform cryogenic photoluminescence on the designed trilayer heterostructure along with time-resolved pulsed measurements via time-correlated single-photon counting, mapping the spectral, pump fluence and bath temperature dependence of the interlayer excitons. We observe an order-of-magnitude larger photoluminescence in the trilayer compared to bilayer heterostructures, due to the more distributed hybridized valance band state in the trilayer for increased oscillator strength, along with the formation of the neutral interlayer exciton. The additional $WSe_2$ layer absorbance in the trilayer also contributes to the stronger photoluminescence compared with the bilayer. At 4K we observe a spectrally-distinguishable trion formation. We determine that the energy splitting between the neutral exciton and trion is largely dependent on the pump fluence and photocarrier densities instead of the bath temperature. We observe that the trion mostly red-shifts while the neutral exciton stays consistent and extract a trion binding energy of approximately 27 meV in our trilayer heterostructure. This result is closed to the interlayer negative charged exciton binding energy of 28 meV predicted by the recent theoretical study on the vertical stack $MoS_2/WS_2$ heterostructures [46]. With increasing photocarrier injection and decreasing bath temperatures, the trion-to-neutral intensity ratio is also increased. Radiative lifetimes of the interlayer excitons are measured up to ≈ 2.54 ns at 4K. We examine the radiative lifetimes for different bath temperatures, pump fluences and excitation wavelengths, with observed stronger dependences on the bath temperature. Our study enables the engineering of indirect neutral excitons and charged trions with spatially separated electron and hole wavefunctions, towards control of the oscillator strengths and excitonic character with designed atomic granularity.

**MATERIALS AND METHODS**

**Computational method:**

In the computational modeling of the $MoSe_2$ and $WSe_2$ based heterostructures, the atomic structure of monolayer $MoSe_2$ and $WSe_2$ are first optimized using first-principles pseudopotential calculations based on the spin-polarized density functional theory (DFT) within the generalized gradient approximation (GGA). The optimized monolayers are placed on top of each other with proper interlayer distances that correspond to the ground state energy for each heterostructure. During this process, van der Waals corrections are implemented following the DFT-D2 method of Grimme [47]. Projector-augmented wave potentials (PAW) [48] are used in the calculations and the exchange-correlation potential is approximated with Perdew-Burke-Ernzerhof functional [48]. During the geometrical optimization steps, the Brillouin zone (BZ) is sampled in the Monkhorst-Pack scheme, where the convergence in energy as a function of the number of $k$-points is tested. The $k$-point sampling of (21×21×1) is found to be suitable for the BZ corresponding to the primitive $MoSe_2$ and $WSe_2$ unit cells. The heterostructures are optimized using the conjugate gradient method, where the total energy and atomic forces are minimized. The energy convergence value between two consecutive steps is chosen as $10^{-5}$ eV. A maximum force of 0.01 eV/ Å is allowed on each atom. In order to avoid interactions between periodic unit cells, a supercell with 20 Å vacuum is used. The optimized heterostructures are then used for the electronic band structure calculations. The band diagrams are obtained by calculating the energies at 150 $k$-points between the Γ-M-K-Γ high symmetry points in the BZ. The effects of spin-orbit coupling on the electronic structures are calculated by performing non-collinear computations as implemented in VASP [50]. The real-space wavefunctions are obtained from the DFT calculations using the binary wavefunction file that is produced by VASP. The binary WAVECAR file that is produced after a series of DFT calculations at the K-point is then mapped by WaveTrans, which correlates the binary wavefunctions with the corresponding lattice points in the reciprocal space and their associated planewave coefficients. The real-space pseudo-wavefunction is constructed again using WaveTrans [51].

**Sample preparation and characterization:**

Single-layer (SL) $MoSe_2$ and SL $WSe_2$ flakes are separately prepared by mechanical exfoliation on clean 290 nm $SiO_2$/Si substrates. A thin layer of hexagonal boron nitride is first exfoliated on a PMMA/PPC stack, which is then used for layer-by-layer picking up of the prepared $WSe_2$ and $MoSe_2$ in order. The picked up $MoSe_2$/$WSe_2$/$h$-BN/polymer stack is finally transferred on a $WSe_2$/$SiO_2$/Si substrate [25,26]. During the entire dry transfer process, no solvent is involved to ensure the atomically clean interface between the 2D materials. Raman measurement (Renishaw, inVia basis) is conducted using a 532 nm laser to map out the sample quality.

**Cryogenic micro-photoluminescence and time-resolved photoluminescence:**

We use an optical cryostat (Janis Research ST-100) to measure the cryogenic photoluminescence at 4 K. Measurements are carried out using a Si detector (Newport Model 2151) for the shorter wavelength regime (400 nm to 1000 nm) with a continuous-wave He-Ne 632.8 nm pump laser. For the longer wavelengths (900 nm to 1600 nm) where interlayer exciton and trion can exist, an InGaAs focal plane array detector with liquid nitrogen cooling system (Princeton Instruments 2D-OMA) is used with high responsivity. The detector signal is improved by using a lock-in amplifier and selectively filtered by a spectrometer (Princeton Instruments SpectraPro 2500i). TRPL measurements are performed using a time-correlated single-photon counting (TCSPC) system with PicoQuant. The sample is excited with a 30 ps 39/78 MHz tunable pulse laser system (NKT Photonics SuperK EXTREME EXW-12) and detected with a single-photon counting module (SPCM-AQR-16) attached to a monochromator and processed by a PicoHarp 300 correlating system with ≈ 300 ps resolution. The TRPL measurements are conducted at low temperature with optical filters in an enclosed optical setup to prevent pump and stray ambient light from affecting the detectors.

**RESULTS AND DISCUSSION**

**Theoretical analysis:**

Here we introduce the concept towards the enhancement of the exciton recombination rate of the trilayer heterostructure illustrated in Figure 1a-b. The trilayer heterostructure consists of a single-layer (SL) $MoSe_2$ sandwiched between two SLs of $WSe_2$. The anticipated type-II band alignment, depicted in Figure 1b, supports excitons with the electron and hole residing respectively in the conduction band of $MoSe_2$ and valence band of $WSe_2$. This resembles the situation for excitons in heterostructure bilayers, where the electron and hole are strongly localized to the different layers by the large band offsets [52-54]. This is indeed the case for the electron in the conduction band of the trilayer structure, which is strongly localized in the central $MoSe_2$ layer due to a weak interlayer coupling [55]. However, in our case, the valence band states are subject to a stronger interlayer coupling [55]. Consequently, the valence band states in our two outer $WSe_2$ layers hybridize with the state in the central $MoSe_2$ layer. Accordingly, the delocalization of the valence band state over the entire trilayer enables the strong overlap of the electron-hole in the trilayer, compared to the prior bilayer counterparts.

To support this physical picture, we carried out first-principles density functional theory (DFT) calculations including the spin-orbit interaction of the bilayer and trilayer band structures and wavefunctions, with the computational details described in the above Methods. The results are summarized in Figures 1c and 1d. (Band structures of SL MoSe$_2$ and SL WSe$_2$ are shown in Supplementary Section S1.) While the bilayer heterostructure shows a ≈ 1.22 eV bandgap, the trilayer heterostructure has a slightly smaller ≈ 1.21 eV bandgap, both at the K-point. In both the bilayer and trilayer structures, our calculations predict states which are predominantly localized in either WSe$_2$ (blue) or MoSe$_2$ (red) layers, in agreement with the band alignment illustrated in Figure 1b. Figure 1e and 1f show the spatial distribution profile of the electron and hole probability densities in the out-of-plane direction, for the states at the valence and conduction band edges. In the bilayer the electron and hole states hardly overlap, hence the long lifetime [31,32,56,57], and our weaker PL shown in Figure S3. On the other hand, in the trilayer structure the valence band state shows the expected delocalization with a large center-layer component and thus a significant overlap with the localized conduction band state.

In our proposed scheme as shown in Figure 1b, the type-II staggered heterojunction leads to energetically lower excitonic states due to the band offset between MoSe$_2$ and WSe$_2$. Furthermore, with the trilayer heterostructure, the strong asymmetry between the electron and hole band profile promotes excitons complexes such as trions [2,16,33,39,58]. Unlike bilayer TMD heterostructures, a metallic state through biasing the Fermi level in the conduction or valence band is required in the trilayers to produce imbalance in electron and hole populations and hence increase the likelihood of trions formation [33,39]. The interlayer trions can be achieved, for example, with two electrons in the MoSe$_2$ layers and one hole in the WSe$_2$ layer. A description of the radiative recombination pathways in the trilayer and the trions is shown in Supplementary Section S2.

**Experimental results and discussion:**

We fabricate the WSe$_2$-MoSe$_2$-WSe$_2$ trilayer heterostructure on a SiO$_2$/Si substrate by dry transfer with a hexagonal boron nitride (*h*-BN) flake of 20 nm thickness as shown in Figure 2a. Both SL WSe$_2$ and SL MoSe$_2$ are mechanically exfoliated from bulk crystals and identified with optical microscopy, PL and Raman spectra. Before the stacking process, the sample thickness is further confirmed independently via atomic force microscopy (AFM) as shown in Figure 2b. Raman spectroscopy is conducted on the SL MoSe$_2$, SL WSe$_2$, 20 nm-thick *h*-BN as well as TMDs trilayer heterostructures as illustrated in Figure 2c. In the inset of Figure 2c, we observe two distinguishable modes at 243 cm$^{-1}$ (MoSe$_2$) and 247 cm$^{-1}$ (WSe$_2$). 20 nm-thick *h*-BN is used as an encapsulating layer due to its atomic smoothness and its relatively absence of charge traps and dangling bonds [59]. Raman shift of *h*-BN shows up in the TMDs-based trilayer heterostructure. The completed *h*-BN/WSe$_2$/MoSe$_2$/WSe$_2$ stacked structure is on a 290 nm SiO$_2$ substrate. To avoid undesired doping, we do not use any chemicals. With regards to the stacking geometry, we randomly stack our TMDs-based vdWs heterostructure since interlayer excitons can potentially be formed regardless of the layer orientation in vertically stacked heterostructures [8]. As shown in Figure 2d, PL spectroscopy is performed on our SL MoSe$_2$ and SL WSe$_2$ before and after transfer at varying temperatures, with the 1.88 eV pump excitation at ≈ 2 mW and with an ≈ 1 μm beam spot size. PL spectra show each peak of the SL MoSe$_2$ (≈ 1.57 eV) and SL WSe$_2$ (≈ 1.64 eV) are obtained with high quantum efficiency; our trilayer shows two PL peaks captured by a Si detector. With decreasing temperature, both MoSe$_2$ and WSe$_2$ peaks are blue-shifted, consistent with prior studies [20,60-62]. The interlayer optical transition from the type-II heterojunction becomes dominant at low temperatures. We noticed from the 77 K and 4 K trilayer PL curves in Figure 2d that when the interlayer PL is more dominated, the intralayer PL is more quenched suggesting an efficient ultrafast charge transfer in the trilayer heterostructure which agrees with previous ultrafast charge transfer studies of the bilayer heterostructure [53, 63-64]. As shown in the light brown line of Figure 2d, we observe that the PL intensities of the interlayer excitons can be larger even compared to those of intralayer excitons. (An example comparison of the trilayer to the bilayer is shown in Supplementary Section S3.) It is interesting to note that the addition of a WSe$_2$ layer to the MoSe$_2$-WSe$_2$ bilayer can result in higher PL quantum yields of both intralayer and interlayer excitons at room temperature and cryogenic temperatures. The reason, as noted in the introduction and theoretical analysis section, is from the strong wavefunction overlap in the trilayer and additional exciton formation from the extra WSe$_2$ layer absorbance. We note that the PL enhancement due to absorbance from the extra layer is not more than a factor of 2, while the enhancement measured is an order-of-magnitude larger. This additional enhancement is the contribution of the increased wavefunction overlap.

Interlayer excitons in coupled quantum wells have previously been studied at low temperature in order to avoid thermal quenching of the PL due to nonradiative processes and electron-hole pairs excited to energies outside the light cone [29,30]. In Figure 3 we illustrate the temperature-dependent PL measurements from 4 K to 245 K of our trilayer atomic crystal, in order to examine the interlayer many-body excitonic transitions. Here we use a liquid-nitrogen-cooled InGaAs photodetector with high responsivity from 900 nm to 1600 nm. In our trilayer heterostructure, a long lower-energy spectral tail can be observed in the main interlayer exciton PL peak and becomes spectrally distinguishable from the neutral interlayer exciton (IEX$_1$) peak at 4 K as shown in Figure 3a (1.96 eV pump excitation). We consequently deconvolved our trilayer interlayer lineshape into two peaks with Lorentzian lineshapes. The higher energy (black line fit) and lower energy (red line fit) peaks are denoted as IEX$_1$ and IEX$_2$ respectively, which we will subsequently attribute to the interlayer neutral excitons and trions (charged exciton) respectively as we will further elaborate below. The inset of Figure 3a shows the spectroscopy with our one-dimensional focal plane array

InGaAs detector, with the horizontal axis the spectra and the vertical axis the spatial position across the sample via a scanning mirror.

Figure 3b shows the interlayer exciton spectra at different bath temperature, with the black ($IEX_1$) and red ($IEX_2$) lines tracking the spectral decomposition. It is worth noting that $IEX_2$ becomes distinguishable only at fairly low temperatures in the interlayer exciton lineshape. As shown in Figure 3b, both peaks are blue-shifted and broadened as the temperature increases and become undetectable above 245 K [20,60-62]. Figure 3c summarizes the measured temperature dependence of their peak positions, quantified by the band gap variation with temperature [62]. Linewidth temperature dependence and the phonon-induced bandgap renormalization model is described in Supplementary Sections S4 and S5 respectively. Under the same pump fluence and as shown in Figure 3c, we note that the energy difference between $IEX_1$ and $IEX_2$ does not change as temperature increases up to 126 K, the temperature after which the $IEX_2$ peak is too weak to distinguish with certainty. This is consistent with the fact that peak splitting of exciton and trion depends on the pumping fluence [16,66], further supported by our power-dependent PL spectra results later in this study.

Figure 3d and 3e show the neutral exciton and trion PL peak intensities and their ratio as a function of temperature. The PL peak intensities decrease with increasing temperature because in, for example, SL $MoS_2$ it is attributed to thermally excited excitons which escape the light cone [67]. Our reduced PL intensity with increasing temperature is attributed to the band structure of our trilayer heterostructure, which resembles that of SL $MoSe_2$ near the band edges despite the existence of two $WSe_2$ layers. In addition we note that the phonon-assisted transitions move the weight from the main PL peak to lower energies and thereby reduce the PL peak intensities. Furthermore, since the promoted nonradiative process competes readily with radiative recombination of the interlayer excitons, the interlayer radiative recombination is not observed at room temperature. In Figure 3e we plot the ratio of PL peak intensities, $I_{IEX2}/I_{IEX1}$, which increases from $\approx 0.13$ at 96 K to $\approx 0.28$ at 4 K due to suppressed nonradiative recombination and high carrier densities. An abrupt rise around 10 K is typically induced by a rapidly increasing lifetime of trions and is a key feature of trions [20]. Discussion about this using mass-action model is in the Supplementary S9. We note that the interlayer radiative recombination channels – in both the neutral exciton and trion – may involve the processes of optical phonons in order to conserve momentum in the band-to-band transition by optical phonon absorption and phonon emission. This is especially the case for the trion which has approximately two times larger FWHM than the neutral exciton (detailed in Supplementary Figure S4) and has a stronger thermal dependence.

In our experimental measurements we have assigned the two distinct interlayer PL emission peaks $IEX_1$ and $IEX_2$ to the neutral exciton and the three-body trion respectively. More details on the interlayer excitonic radiative recombination pathways are described in Supplementary Section 2 and Figure S2. For the lower energy ($IEX_2$) interlayer peak, we have also considered two other scenarios but both are unlikely. The first is a phonon-assisted indirect radiative recombination between valence band maximum (VBM) of $WSe_2$ at K-point and conduction band minimum (CBM) of $MoSe_2$ at $\Sigma$-point, per the band diagram of Figure 1c and 1d. However, based on our DFT calculations, the second valley at $\Sigma$-point of $MoSe_2$ CBM is located at $\approx 200$ meV higher than K-point of $MoSe_2$ CBM. This large energy requirement makes it hard for the carriers to thermalize in this phonon-assisted transition, even at room temperature ($\approx 25.7$ meV). The second considered scenario is that of a potential sub-band above the K-point of $MoSe_2$ CBM. However, in our DFT calculations, we could not find the sub-band or energy splitting at $MoSe_2$ CBM nor the $WSe_2$ VBM. This further supports that the $IEX_2$ transition arises from the interlayer trions.

In Figure 4 we summarize the pump fluence dependence of the two interlayer excitons in our trilayer heterostructure by tuning the optical injection of the carriers. While the spectral shift of interlayer neutral exciton under increasing pump fluence has only a slight blue-shift, the spectral position of the trion is significantly more red-shifted with increasing pump fluence. This is illustrated in Figure 4a. Figure 4b summarizes the extracted energy difference between the peaks of the two electronic states for increasing photocarrier injection, illustrating a resulting linear dependence. A description of the two peak energy difference and the trion binding energy, in the presence of carrier screening, is detailed in Supplementary Section S2. Based on this energy splitting, we extract the trion binding energy of $\approx 27$ meV from Figure 4b [19], which is higher than the thermal activation energy at room temperature and even comparable with intralayer trion binding energy of SL TMDs [16,19-21,58,68], and $\approx 28$ meV negative charged interlayer exciton binding energy in vertical stack MoS2/WS2 heterostructures [46]. This increasing spectral shift and binding energy with photocarrier density is another evidence for the interlayer trions [19,39]. This is because, with increasing photocarrier injection, free charge requires more energy to occupy the three-body trion excitonic states. In turn, we plot the PL spectral weights of the neutral exciton and trion in Figure 4c for increasing excitation fluence. With increasing photocarrier injection, the spectral weight and the ratio of the trion to the neutral exciton increases, which is supporting evidence of the interlayer trions existence [39].

We further investigate the lifetimes of the interlayer neutral exciton and trion, to help elucidate the many-body process. We build the TRPL setup by means of a time-correlated single-photon counting system (TCSPC) and conduct temperature-, pump fluence-, and pump wavelength-dependent TRPL experiments. All results are fitted biexponentially to the experimental data. Radiative recombination of excitons is difficult to obtain at room temperature because nonradiative recombination is dominant. Accordingly we perform lifetime measurements at cryogenic temperatures, with ≈ 9 µJ/cm$^2$ pump fluence and 1.88 eV (660 nm) pump wavelength, above both the SL MoSe$_2$ and SL WSe$_2$ direct gaps. Figure 5a and 5b summarizes the recombination lifetimes of interlayer neutral exciton and trion at various cryogenic temperatures. We note that, in the measurement of the IEX$_2$ lifetimes, we band-pass filter the spectrum with center wavelength around 990 ± 2 nm and FWHM of 10 ± 2 nm, to select the PL centered ≈ 985 nm in order to exclude the dynamics of the interlayer neutral exciton. As shown in Figure 5a, the lifetimes of our trilayer heterostructure interlayer exciton (IEX$_1$) are ≈ 2.54 ns (4 K), 2.47 ns (40 K), 1.84 ns (80 K), and 0.41 ns (120 K), on the same order-of-magnitude as prior bilayer heterostructure studies [52] and theoretical calculations [32]. For our trilayer heterostructure interlayer trion (IEX$_2$) they are ≈ 1.24 ns (4 K), and 0.47 ns (20 K). At higher bath temperatures, nonradiative recombination (including phonon-assisted processes) is faster, resulting in the shorter effective lifetimes observed.

Figure 5c summarizes the extracted lifetimes from Figure 5a and 5b. Although the data points are sparse, we note that the lifetime of the interlayer excitons decreases linearly with increasing temperature and the interlayer trions decreases faster than interlayer excitons. This also correlates with the observed faster trion lifetimes in the finite bath temperature TRPL and the broader linewidths of the trions. We note that the lifetime of the interlayer trion is not observable above 20 K due to the low photon counts. To further explore the properties of interlayer exciton, we turn our attention to the interlayer exciton lifetimes under various pump fluences and pump wavelengths. As shown in Figure 5d and 5e, the lifetime of interlayer exciton (IEX$_1$) is weakly dependent on carrier injection compared to the bath temperature. As the pump fluence increases, the lifetime only decreases slightly from ≈ 2.72 ns (1.6 µJ/cm$^2$) to ≈ 2.53 ns (8.1 µJ/cm$^2$). With increasing pump fluence, the interlayer exciton density increases which causes a blue-shift. The dipolar exciton-exciton interaction may lead to the change of peak positions and the decrease of lifetime [69]. The decrease of lifetime with increasing excitation fluence indicates that residual defect states are involved, as shown in Figure 5d for the IEX$_1$ interlayer lifetime (lifetime of IEX$_2$ with increasing excitation fluence is not distinctly resolvable). The presence of residual defects also explains why the PL intensities of interlayer exciton and trion decrease as temperature increases. We also conduct pump wavelength-resolved TRPL experiment which is detailed in Supplementary Section S6. There is no correlation between the pump wavelength and the lifetimes of the interlayer exciton. This is because the electron-phonon decay dynamics to the band edge in TMDs is on the order of picoseconds, much shorter than the radiative recombination of interlayer excitons [70-72].

## CONCLUSIONS

In this study we demonstrated the enhanced Coulomb interactions in $WSe_2$-$MoSe_2$-$WSe_2$ trilayer vdWs heterostructures, through interlayer radiative recombination of the neutral and charge excitonic states. Designed through spin-polarized density functional theory with exchange correlation and spin-orbit coupling, we examined type-II heterostructures with increased optical transition oscillator strengths. Fabricated via $h$-BN enabled dry transfer, the trilayer heterostructure exhibits an ≈ 3× (at 77 K), and 5× (at lower temperature) stronger photoluminescence in our measurements. This arises from the more spatially distributed hybridized valence band state across the trilayer which, while the conduction band is localized only at the $MoSe_2$ layer, enables larger overlap and oscillator strengths. Our cryogenic steady-state and time-resolved photoluminescence measurements elucidate the additional presence of a three-body excitonic state in the interlayer radiative recombination, with the trion binding at ≈ 27 meV and largely dependent on the pump fluence and carrier densities instead of the bath temperature. Conversely the trilayer interlayer radiative lifetimes are largely dependent on the bath temperature instead of the pump fluence and exciton wavelengths, promoted by phonon-assisted transitions. The relatively long radiative lifetimes measured up to 2.54 ns at 4 K enables next-generation excitonic devices [73]. Future experiments include gating the heterostructures with controlled carrier densities and spectrally separated excitonic resonances to further clarify their roles in exciton dephasing and population relaxation [17, 74]. Also, previous studies about the ultrafast charge transfer between layers within the $MoS_2$/$WS_2$ heterostructures reveal the fs scale dynamics which can be tuned significantly by changing stacking configuration [53, 63-64]. So future ultrafast pump-probe spectroscopy experiment on our trilayer compared with the bilayer heterostructure can be an interesting topic about the effect of extra $WSe_2$ layer, inducing its stacking configuration, on the charge transfer dynamics. With advanced 2D material growth techniques and computational methods [27,75], we believe our trilayer heterostructures support the understanding of interlayer light-matter interactions with facilitated carrier transfer and type-II heterojunctions, paving new ways to tailor the electronic and optoelectronic character of vdWs heterostructures.

**Author contributions:** C.C. designed this work. C.C., H.K. and J.H. performed the optical measurements and data analysis. H.-C.C. and C.C. led the device fabrication. C.C., J.C, S.-H.B. and X.D. provided TMDs materials and sample characterizations. C.C., A.K.V., S.-W.H., V.O.Ö., R.G., K.K., T.L and C.W.W. contributed the theoretical analysis and simulations. C.W.W. supported and supervised the research. C.C., K.K., T.L. and C.W.W. prepared the manuscript.

**Funding sources:** We acknowledge support from the University of California – National Laboratory research program and the National Science Foundation (DMR-1611598).

**Financial interest:** The authors declare no competing financial interest.

**Acknowledgements:** The authors are grateful for helpful discussions with Philip Kim, Tony van Buuren, Baicheng Yao, Jinkang Lim, Yandong Luo, Zhangji Zhao, and Jin Ho Kang.

**Supplementary Information:** Supplementary Information for this article can be found online.

**Data Availability:** All relevant data are available upon the request from corresponding authors.

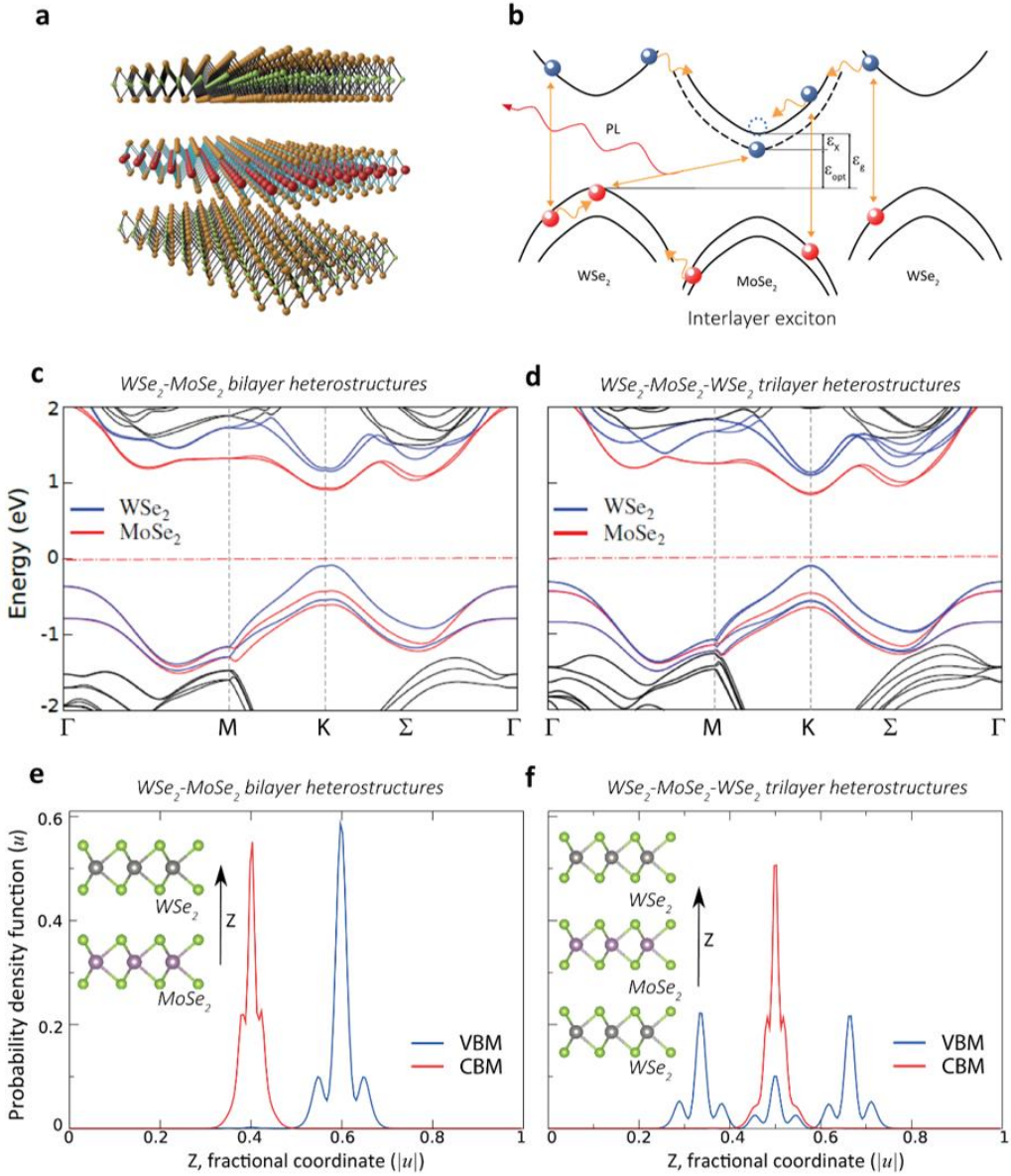

**Figure 1 | Excitonic states of TMDs-based trilayer heterostructure, interlayer excitons model, and computed band structure and wavefunction distribution comparisons between the bilayer and trilayer heterostructures. a,** Illustration of $WSe_2$-$MoSe_2$-$WSe_2$ trilayer heterostructure with a sandwiched single layer (SL) $MoSe_2$ between two SLs $WSe_2$. **b,** Schematic summary of the interlayer exciton radiative recombination predicted by type-II heterojunction. $\varepsilon_X$, $\varepsilon_{opt}$ and $\varepsilon_g$ represent the exciton binding energy, optical gap and electronic gap respectively. **c,** Computed band structure of the $WSe_2$-$MoSe_2$ bilayer heterostructure. **d,** Computed band structure of the $WSe_2$-$MoSe_2$-$WSe_2$ trilayer heterostructure. **e,** Computed orbital wavefunction of the $WSe_2$-$MoSe_2$ bilayer heterostructure valence band maximum (VBM) and conduction band minimum (CBM) at *k*-point as a function of the interlayer distance *z*. **f,** Computed orbital wavefunction of the $WSe_2$-$MoSe_2$-$WSe_2$ trilayer heterostructure.

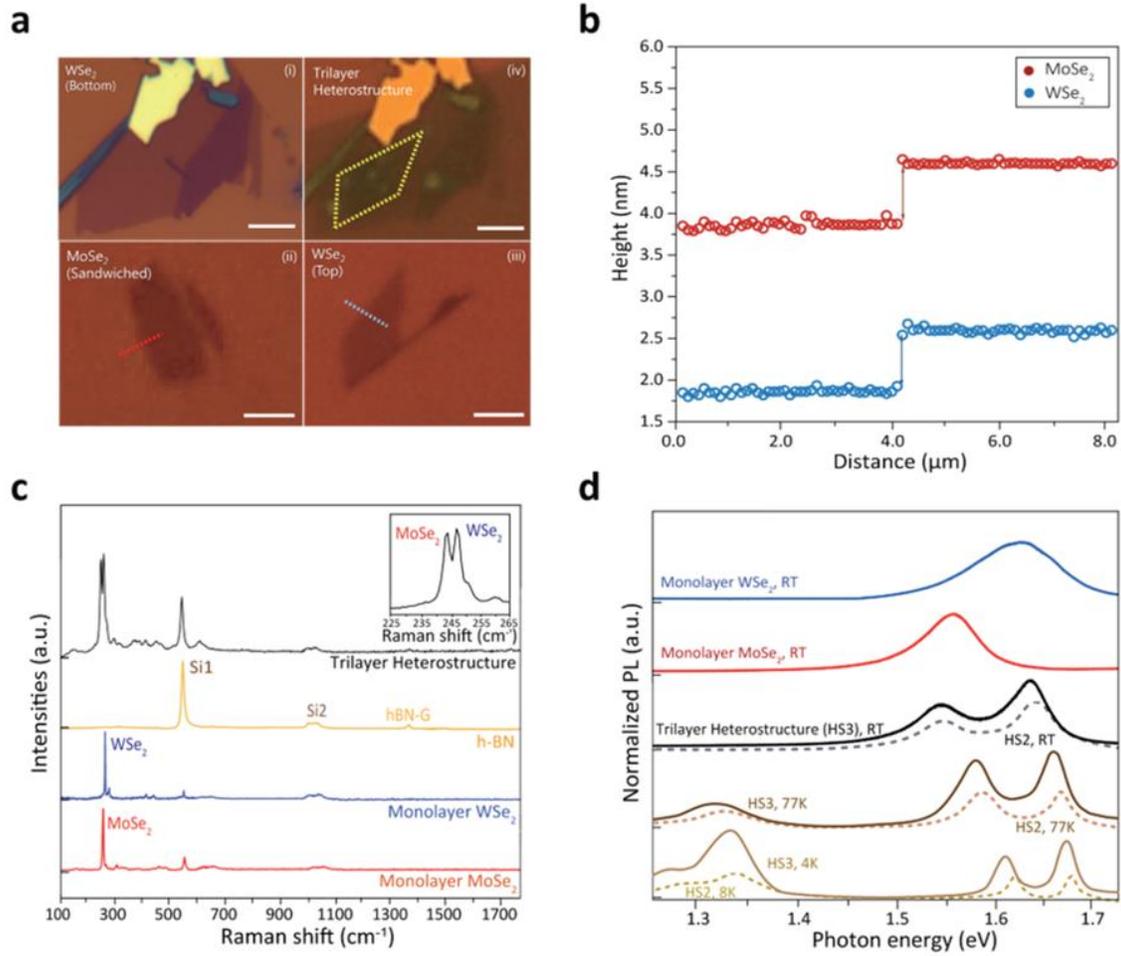

**Figure 2 | TMDs-based trilayer van der Waals (vdWs) heterostructure fabrication and characterization. a,** Microscopic images of the exfoliated transition metal dichalcogenides monolayers [i – iii] and trilayer vdWs heterostructure [iv] on $SiO_2$ substrate. The dashed red and blue lines in the lower panels denote the atomic force microscopy (AFM) regions. The area enclosed by dashed yellow line indicates the trilayer heterostructure region. Scale bar: 5 μm. **b,** AFM cross-sectional profiles, where each atomic single-layer (SL) is ≈ 7Å. $WSe_2$ and $MoSe_2$ are the blue and red lines respectively. **c,** Raman spectra of the trilayer heterostructure versus each SL ($MoSe_2$, $WSe_2$, $h$-BN) decomposition, pumped with a 532 nm laser. Inset: zoom-in of the Raman shift around 245 $cm^{-1}$. **d,** Room-temperature PL spectra for the SL $WSe_2$, SL $MoSe_2$ and the trilayer heterostructure, together with the low-temperature PL spectra from the trilayer heterostructure at 77 K and 4 K. The excitation is via a 632 nm HeNe laser. Dotted lines under the trilayer heterostructure PL spectra are the bilayer heterostructure PL spectra, with especially large enhancement of the lower-energy interlayer exciton transition.

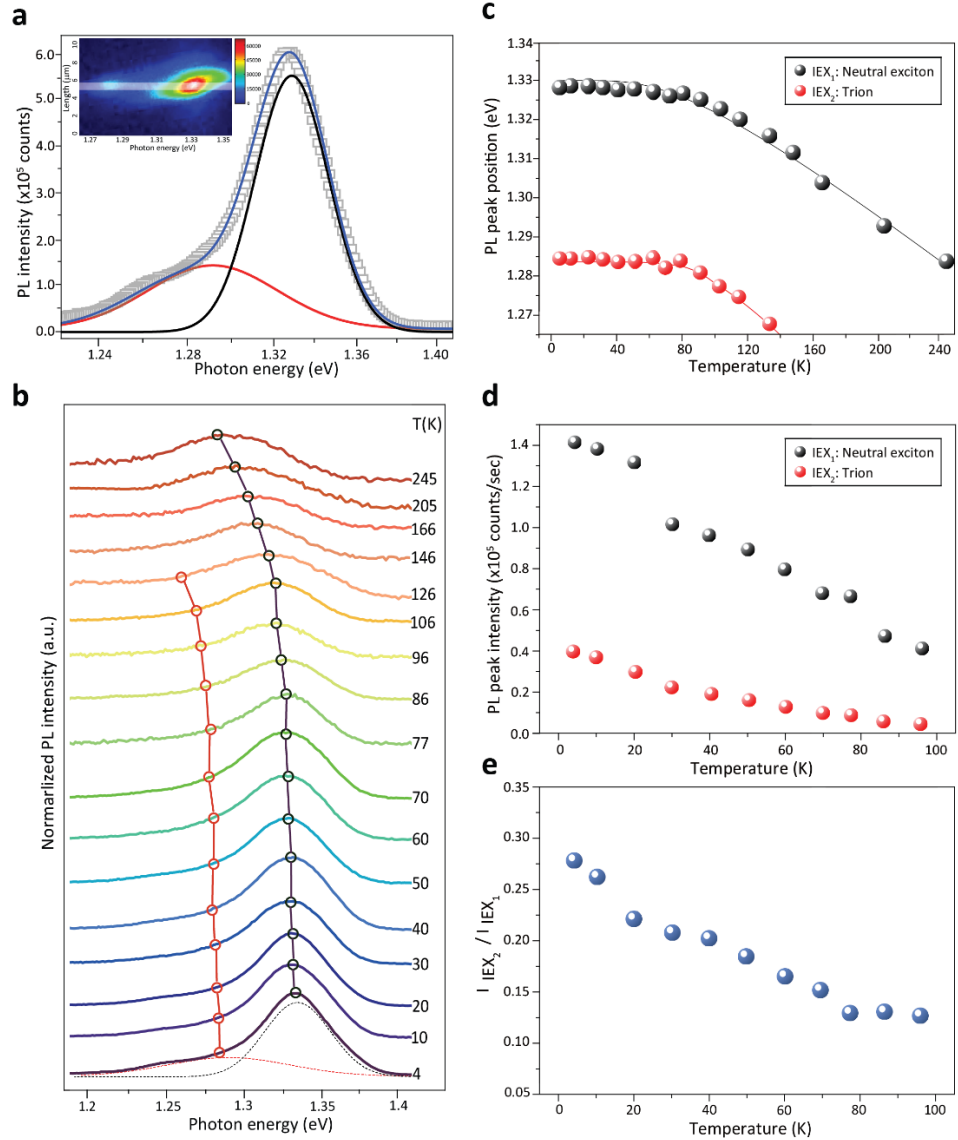

**Figure 3 | Temperature-dependent PL intensities and spectral lineshapes of the interlayer excitons in the trilayer heterostructure. a,** Indirect excitons PL spectra at 4 K, fitted with a blue line. Both IEX$_1$: neutral exciton (black) and IEX$_2$: trion (red) lines are fitted with Lorentzian lineshapes. Inset image shows PL spectrum captured by a focal plane array detector with horizontal axis the spectra and the vertical axis the spatial position across the sample in one direction. **b,** Temperature-dependent indirect PL spectra in trilayer heterostructure with pump excitation at 1.96 eV and ≈ 1.2 mW reaching sample. The black and red lines denoted the respective peaks, with the dashed black and red curves at the 4K spectra illustrating the Lorentzian lineshape decomposition. The data were obtained from the same spot on the sample. The spectra are offset for clarity. **c,** PL peak shifts of the neutral exciton (IEX$_1$) and trion (IEX$_2$) deduced from the main panel and fitted with semiconductor bandgap temperature-dependence model. The phonon-induced bandgap

renormalization is described in Supplementary Section S5. **d,** PL peak intensities at different cryogenic temperatures. Black and red dots indicate the PL intensities of neutral exciton and trion respectively. **e,** The ratio of neutral exciton ($IEX_1$) to trion ($IEX_2$) peak intensities in the trilayer heterostructure are illustrated as a function of bath temperature.

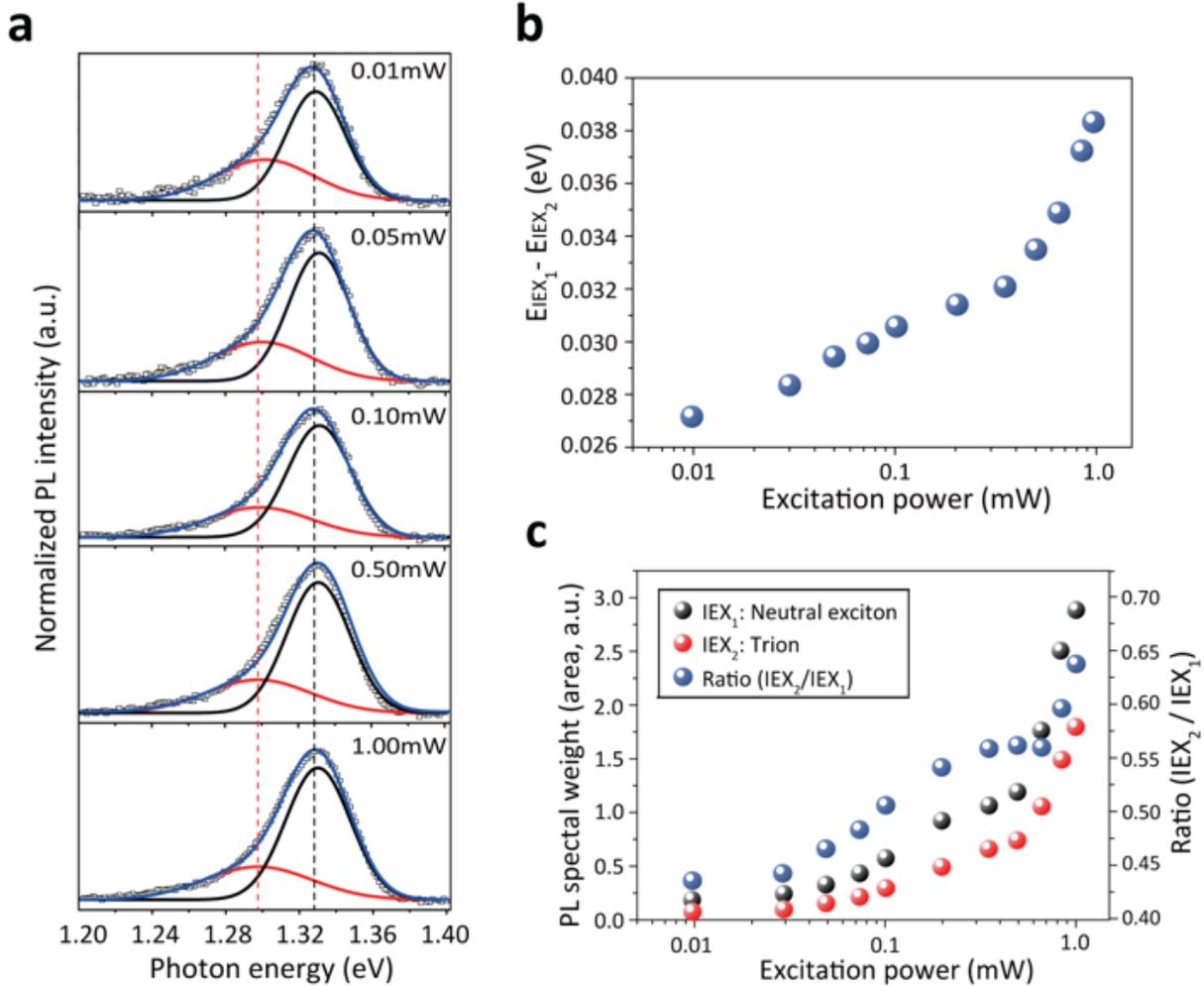

**Figure 4 | Power-dependent PL spectra in the trilayer heterostructure at 4 K. a,** Power-dependent PL spectra with Lorentzian-fitted neutral exciton (black curve) and trion (red curve) at 4 K. **b,** Energy splitting (**t**rion binding energy**:** $E_{IEX1}$ - $E_{IEX2}$) between two interlayer optical transition peaks as a function of excitation power (log scale). **c,** PL spectral weight of neutral exciton (black dots) and trion (red dots), and the ratio of trion to neutral exciton (blue dots), as a function of pump excitation power.

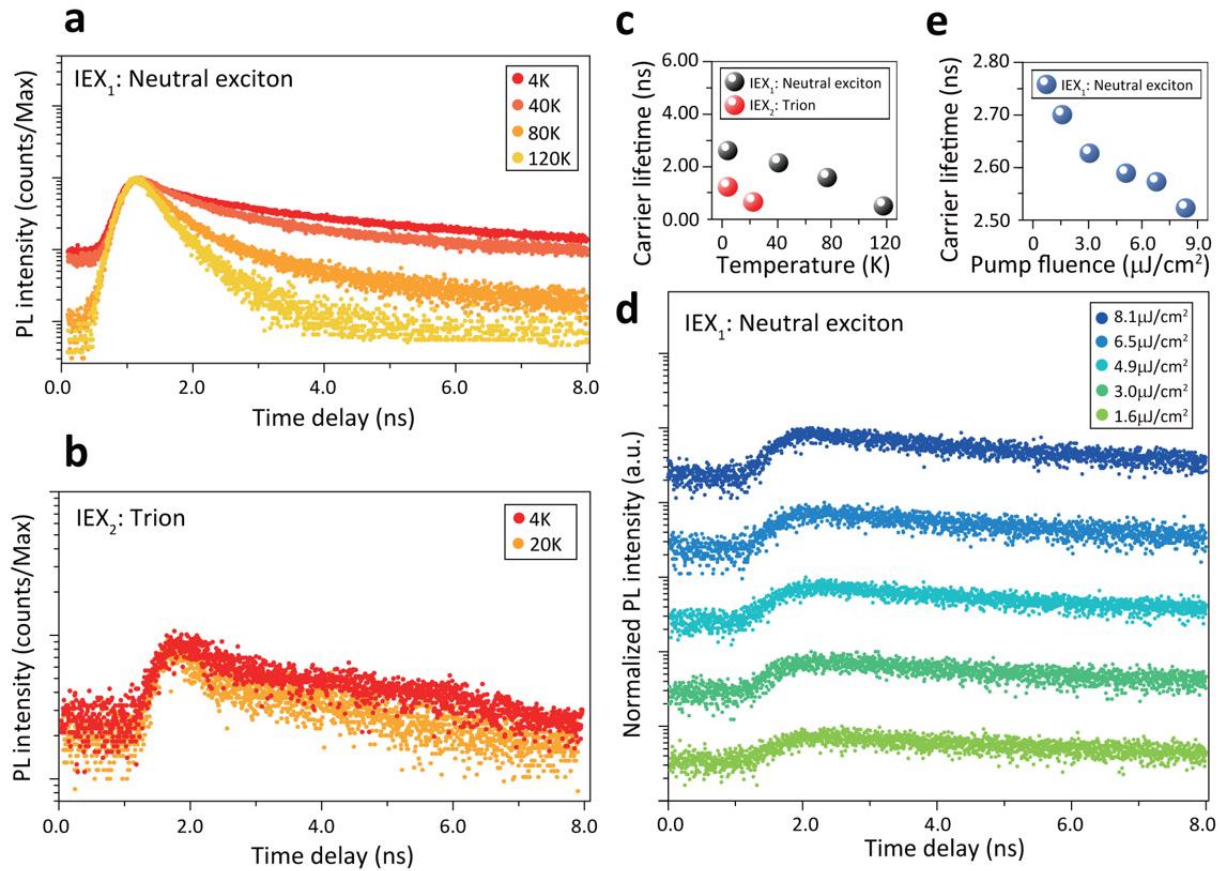

**Figure 5 | Interlayer excitons lifetime and its dependence on temperature and photocarrier injection in the trilayer heterostructure. a** and **b**, Temperature-dependent carrier lifetimes of interlayer neutral exciton ($IEX_1$) and trion ($IEX_2$) in trilayer heterostructure. Samples are excited with ≈ 9 μJ/cm² power fluence at 1.88 eV (660 nm). **c,** Carrier lifetimes of $IEX_1$ and $IEX_2$ with different temperatures at pump fluence of ≈ 9 μJ/cm². The carrier relaxations are fitted with exponentials to extract the carrier lifetimes. **d,** Time-resolved radiative recombination of $IEX_1$ under different pump fluences at 4 K. **e,** Carrier lifetimes of $IEX_1$ under different pump fluences at 4 K.

## Supplementary Information

**Enhanced interlayer neutral excitons and trions in trilayer van der Waals heterostructures**

Chanyeol Choi[1,2,9,†], Jiahui Huang[1,2], Hung-Chieh Cheng[3,4], Hyunseok Kim[2], Abhinav Kumar Vinod[1,2], Sang-Hoon Bae[3,4,10], V. Ongun Özçelik[5], Roberto Grassi[6], Jongjae Chae[3], Shu-Wei Huang[1,2,11], Xiangfeng Duan[4,7], Kristen Kaasbjerg[8], Tony Low[6], and Chee Wei Wong[1,2,†]

[1] Fang Lu Mesoscopic Optics and Quantum Electronics Laboratory, University of California, Los Angeles, CA 90095, United States
[2] Department of Electrical Engineering, University of California, Los Angeles, CA 90095, United States
[3] Department of Materials Science and Engineering, University of California, Los Angeles, CA 90095, United States
[4] California Nanosystems Institute, University of California, Los Angeles, CA 90095, United States
[5] Andlinger Center for Energy and the Environment, Princeton University, Princeton, New Jersey 08544, United States
[6] Department of Electrical and Computer Engineering, University of Minnesota, Minneapolis, MN 55455, United States
[7] Department of Chemistry and Biochemistry, University of California, Los Angeles, CA 90095, United States
[8] Center for Nanostructured Graphene (CNG), Department of Micro- and Nanotechnology (DTU Nanotech), Technical University of Denmark, DK-2800 Kgs. Lyngby, Denmark
[9] Present address: Department of Electrical Engineering and Computer Science, Massachusetts Institute of Technology, Cambridge, MA 02139, United States
[10] Present address: Department of Mechanical Engineering, Massachusetts Institute of Technology, Cambridge, MA 02139, United States
[11] Present address: Department of Electrical, Computer, and Energy Engineering, University of Colorado Boulder, Boulder, CO 80309, United States

[†] Corresponding authors: cowellchoi@gmail.com ; cheewei.wong@ucla.edu

### Content

S1. Computed band structures of single-layer (SL) $MoSe_2$ and $WSe_2$
S2. Two interlayer excitons radiative recombination pathways
S3. Integrated photoluminescence (PL) intensity in bilayer and trilayer heterostructures
S4. Full width half maximum (FWHM) of two interlayer excitonic states
S5. Semiconductor bandgap renormalization with temperature
S6. Time-resolved photoluminescence (TRPL) with different excitation wavelengths
S7. Cryogenic micro-PL measurement of individual $WSe_2$ and $MoSe_2$ monolayer
S8. Cryogenic PL mapping of the trilayer heterostructure
S9. Mass-action model analysis of interlayer exciton trion concentration

## S1. Computed band structures of single-layer (SL) MoSe$_2$ and WSe$_2$

Figure 1 shows the results of computed band structures for SL MoSe$_2$ and WSe$_2$, based on the spin-polarized density functional theory (DFT) within the generalized gradient approximation (GGA) using the same set of parameters as presented in the main text. Both of these SL structures have $D_{3h}$ group symmetry without inversion symmetry and have a direct band gap at the K point. The lattice constants of both MoSe$_2$ and WSe$_2$ SL structures are calculated as 3.33 Å with a nearest neighbor Mo(W) – Se distance of 2.54 Å. In the electronic structure calculations, when the spin-orbit coupling is excluded, the spin-up and spin-down valance bands are degenerate at the K point. However, spin-orbit interaction splits these bands from each other by 0.18 eV and 0.47eV at the K point for MoSe$_2$ and WSe$_2$ respectively. On the other hand, the spin-orbit interaction at the conduction band minimum is much lower, 0.02 eV for MoSe$_2$ and 0.03 eV for WSe$_2$. Here it should be noted that at the K point the valence band maximum is dominated by the $d_{x2-y2}$ and $d_{xy}$ orbitals of the Mo/W atoms and the conduction band minimum is dominated by the $d_{z2}$ orbitals of the Mo/W atom. It should also be noted that the chalcogen atom Se has no contribution to the VBM or CBM at the K point.

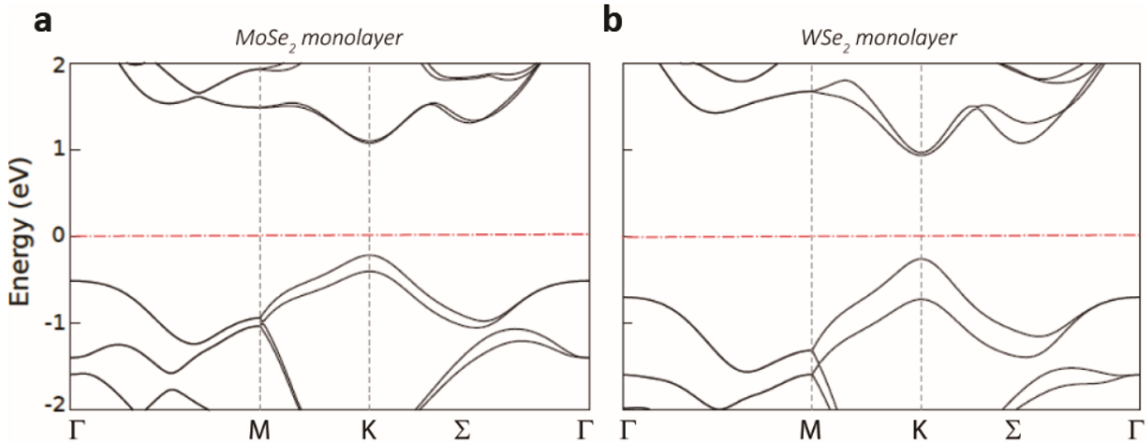

**Figure S1 | Computed band structures of single-layer MoSe$_2$ and WSe$_2$. a,** Computed band structure of SL MoSe$_2$. **b,** Computed band structure of SL WSe$_2$.

## S2. Two interlayer excitons radiative recombination pathways

Figure S2 shows two possible ways to explain a lower energy peak: one is another interlayer exciton (IEX$_2$) and the other is interlayer trion. As for explanation of another interlayer exciton described in Figure S2a, while IEX$_1$ occurs at the same momentum space (K-K transition), IEX$_2$ occurs through pathway placing on different momentum space (K-Σ transition). To make it feasible, the energy variance between the Σ edge at MoSe$_2$ conduction band and K edge at WSe$_2$ valence band should be on the same order-of-magnitude as the energy caused by thermal fluctuation. When it comes to explanation of interlayer trion as illustrated in Figure S2b, trions are generated through interaction between interlayer exciton and electron due to electron accumulation in MoSe$_2$ conduction band minimum on a picosecond timescale [S1,S2]. In effect, we can obtain the trion binding energy ($\varepsilon_T$) through the power-dependent photoluminescence. The tuning effect of three-body excitonic system results from the strong quantum-confined Stark effect [S3]. The energy splitting between interlayer neutral exciton and trion is expressed [S4]:

$$\omega_I - \omega_X = E_{Trion} + \Delta E \tag{S1}$$

where $E_{Trion}$ is the trion binding energy and $\Delta E$ is the energy required for one carrier to be promoted into the free-carrier system. $\omega_I - \omega_X$ denotes the minimum energy for the removal of one electron from the trion, because the exciton is regarded as an ionized trion [S5]. $\Delta E$ becomes negligible when the Fermi surface increases with high carrier injection since the initial doping level ($E_F$) is less significant as a result of the enhanced carrier screening effect by carrier injection in 2D systems [S6]. In our case, we can expect high quantum PL yield, efficient charge transfer, and large effective mass of the carriers in the trilayer heterostructures. Therefore, $\Delta E$ becomes negligible and $\omega_I - \omega_X$ can be taken as the trion binding energy [S4]. In our trilayer heterostructure, trion binding energy is ≈ 27 meV, which deduced from Figure 4b.

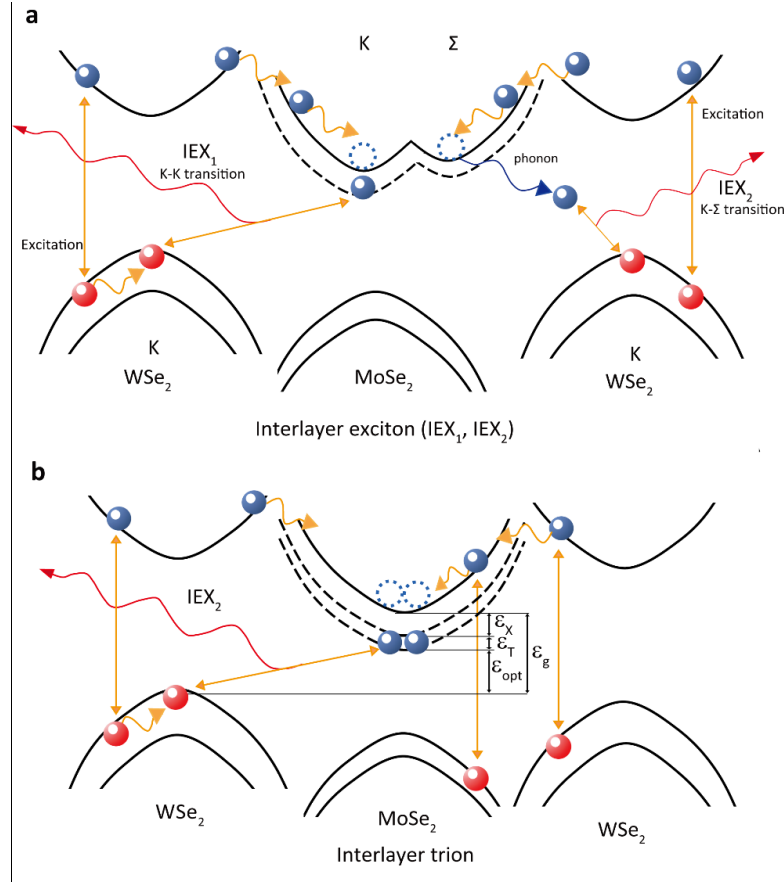

**Figure S2 | Illustration of two pathways for interlayer exciton radiative recombination, and the interlayer trions model. a,** Illustration of two different interlayer excitons radiative recombination pathways: one is K-K direct transition in *k*-space and the other is K-Σ indirect transition in *k*-space. In the case of indirect transitions, phonon is involved to match the momentum difference, which lowers the energy bandgap. **b,** Schematic summary of the interlayer trion

radiative recombination predicted by a type-II heterojunction in our trilayer vdWs heterostructure. $\varepsilon_X$, $\varepsilon_T$, $\varepsilon_{opt}$ and $\varepsilon_g$ represent exciton binding energy, trion binding energy, optical bandgap and electronic bandgap, respectively. This is an example in the formation of the interlayer trions: two electrons in $MoSe_2$ and one hole in $WSe_2$.

**S3. Integrated photoluminescence (PL) intensity in bilayer and trilayer heterostructures**

To compare bilayer heterostructure with trilayer heterostructure, integrated PL intensities are measured at ≈ 0.2 mW and with an ≈ 1 μm laser spot size. As shown below, ≈ 3 times brighter PL intensity (and ≈ 30 meV lower energy PL peak) of trilayer heterostructure is observed at 77 K. To show the PL intensity difference of two different heterostructures, we collected the below data in the same bath temperature and excitation time duration by one dimensional InGaAs focal plane array detector. The noise here is caused by the absence of filters, only for this measurement.

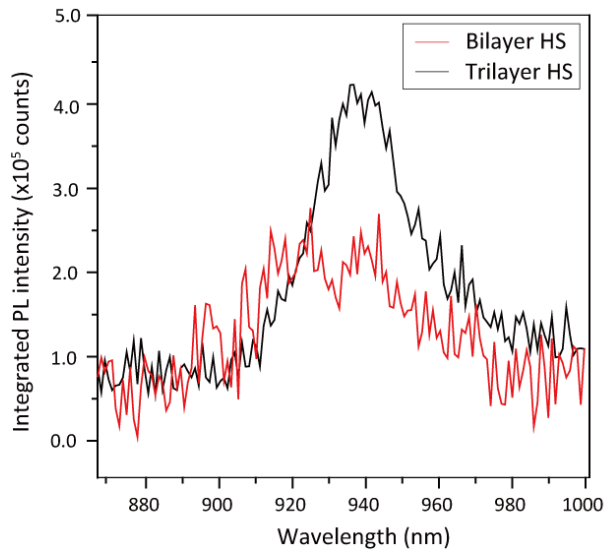

**Figure S3 | Integrated PL intensities in bilayer and trilayer heterostructures.** This comparison demonstrates the brighter PL emission in $WSe_2$-$MoSe_2$-$WSe_2$ trilayer heterostructure than $WSe_2$-$MoSe_2$ bilayer heterostructure, enabled by the tightly overlapping wavefunctions and additional absorbance from the extra $WSe_2$ layer in the trilayer heterostructure.

**S4. Full-width half-maximum (FWHM) of two interlayer excitonic states**

Figure S4 plots the FWHM of the two interlayer excitonic states. In order to see the temperature effects on the interlayer excitonic states, we extrapolated both sets of FWHM data from Figure 3b and note that $IEX_2$ features a much broader FWHM than $IEX_1$. From 4 K to 96 K, we can see that the FWHM of two interlayer excitonic states is almost constant with increasing temperature, with the interlayer trion ($IEX_2$) showing a larger width. The tendency of not broadening with increasing temperature suggests that there are more processes, such as defect scattering or other radiative recombination, contributing to the FWHM other than electron-phonon interactions [S7].

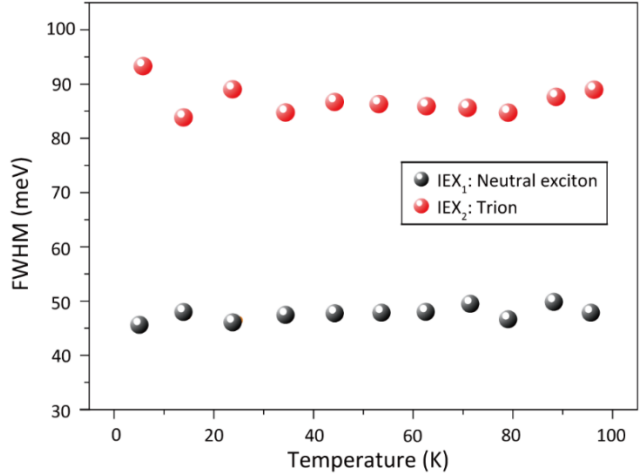

**Figure S4 | Full-width half-maximum (FWHM) of two interlayer excitonic states.** FWHM of interlayer exciton and trion in the trilayer heterostructure as a function of the bath temperatures.

## S5. Semiconductor bandgap renormalization with temperature

In Figure 3c the two PL peaks of excitonic states (neutral exciton and trion) are fitted by temperature-bandgap renormalization model as below [S8]:

$$E_g(T) = E_g(0) - S\langle\hbar\omega\rangle\left[\coth\left(\frac{\langle\hbar\omega\rangle}{2K_b T}\right) - 1\right] \quad (S2)$$

where $E_g(0)$ represents the band gap at zero temperature, $S$ is a dimensionless coupling constant, and $\langle\hbar\omega\rangle$ is an average phonon energy. The above describes the phonon-induced gap reduction with temperature and we fit the neutral exciton peak shift (solid lines in Figure 3c) with corresponding $E_g(0) \approx 1.331$ eV (interlayer neutral exciton ground-state transition energy, 0 K), $S \approx 0.006$ and $\langle\hbar\omega\rangle \approx 6.0$ meV. The trion peak is also fitted by values for $S \approx 0.012$, $\langle\hbar\omega\rangle \approx 9.9$ meV, and $E_g(0) \approx 1.284$ eV (interlayer trion ground-state transition energy, 0 K). The energy splitting between the two excitonic states peaks is $\approx 47$ meV at 4 K, with 1.96 eV pump and 1.2 mW pump power. It decreases with lower pump power and is $\approx 27$ meV at 0.01 mW pump. We focus on the measured data from 4 K to 245 K since above 245 K the PL signals from interlayer neutral exciton and trion become almost indistinguishable from the noise.

## S6. Time-resolved photoluminescence (TRPL) with different excitation wavelengths

In addition to the temperature-dependent lifetime measurements, we also performed TRPL measurements by varying the excitation wavelengths to explore the electron-phonon decay dynamics of the interlayer neutral exciton in the trilayer heterostructure. As shown in Figure S6a, the wavelengths of excitation pulse pump range from 510 nm (green) to 660 nm (red). In each measurement, we selected bandpass filters to exclude the effect of pump side-peaks before our trilayer sample. We further performed TRPL measurements for the interlayer trion; however, its comparably lower photon counts and weaker pump excitation pulse energy below 600 nm does not allow a lifetime to be meaningfully obtained. We also note that our tunable pulse laser has different pulsewidths for different pump excitation wavelengths as shown in Figure S6b. The extracted carrier lifetimes, with a summary shown in Figure S6c, are on the order of nanoseconds and significantly longer than the picosecond pump pulsewidth variations.

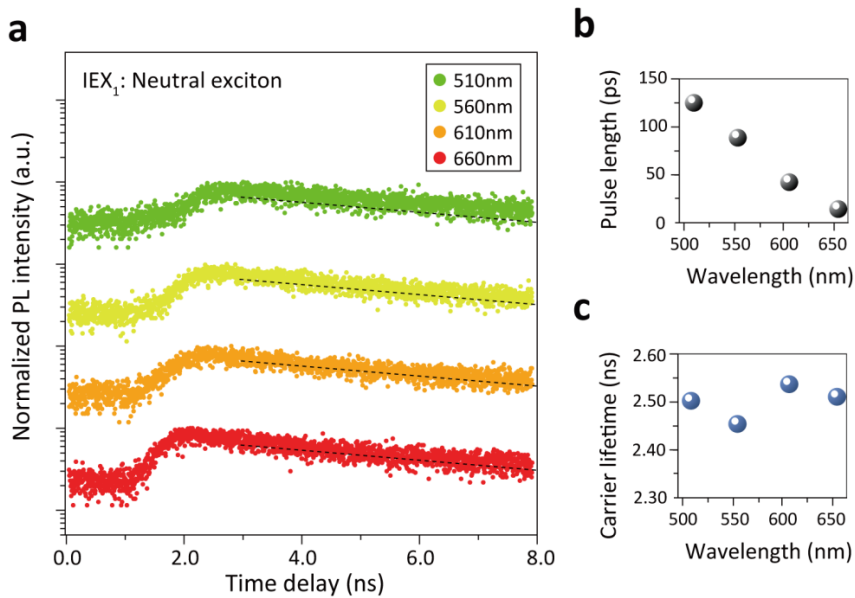

**Figure S6 | Carrier lifetimes for various excitation wavelengths. a,** Interlayer exciton lifetime as a function of excitation wavelengths at 4 K. Photoluminescence intensities are normalized and fitted with biexponential decay functions. **b,** Excitation pulsewidths as function of excitation wavelengths, from the pump laser. **c,** Measured carrier lifetimes deduced from panel **a**.

### S7. Cryogenic micro-PL measurement of individual WSe$_2$ and MoSe$_2$ monolayers

Defect trapped localized exciton recombination may generate localized PL emission at low temperature. Previous studies observed PL emission of localized excitons of monolayer MoSe$_2$ and WSe$_2$ at 1.58 to 1.66 eV and 1.64 to 1.69 eV [S9-S11]. However, the existence of localized emission around 1.28 and 1.33 eV is unknown. To rule out the possibility that the interlayer emission are from the localized exciton, we performed cryogenic micro-PL measurement of the MoSe$_2$ and WSe$_2$ monolayers from the same bulk crystal, examining from 1.18 to 1.58 eV as shown in Figure S7. No localized emission is observed in the 1.28 and 1.33 eV region.

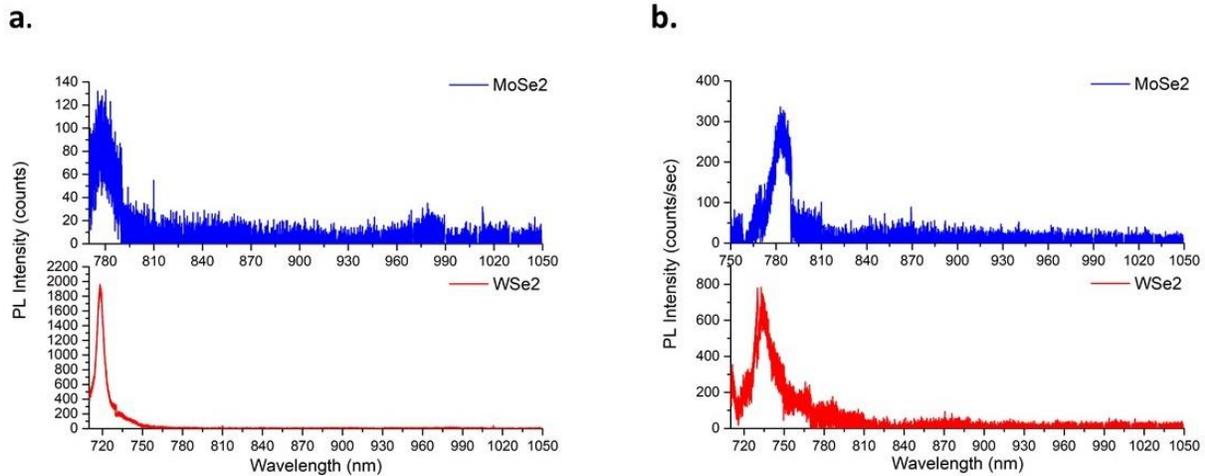

**Figure S7 | Cryogenic micro-PL measurement of individual monolayer MoSe$_2$ and WSe$_2$ from the same bulk. a,** micro-PL measurement of monolayer MoSe$_2$ and WSe$_2$ from the same bulk at 77 K. **b,** micro-PL measurement of monolayer MoSe$_2$ and Wse$_2$ from the same bulk at 4 K. Both measurements are using the 660 nm laser excitation.

### S8. Cryogenic PL mapping of trilayer heterostructure

Previous studies of the strain effect on the stacked van der Waals heterostructures and 2D materials have reported the PL spectral shift and intensity change due to tensile strain and strain-induced effects, like the strain-induced indirect-to-direct bandgap transition [S12, S13]. To clarify the strain effect in the trilayer heterostructure, the PL intensity mapping around the trilayer region at 77 K is performed Only interlayer PL is collected by using the 850 nm long pass filter. As shown in figure S8, the trilayer heterostructures is possessing a uniform PL intensity over the interlayer spectral window. The transition region (yellow region in figure S8) between high and low intensity is due to the program averaging when the laser spot is on the edge of the trilayer flake and slightly out-of-focus of the laser. This can confirm with the expectation that less strain effects, such as bending a substrate during the material transfer processes, are involved or the strain is uniform on the sample so that the strain is not a major impact in our study.

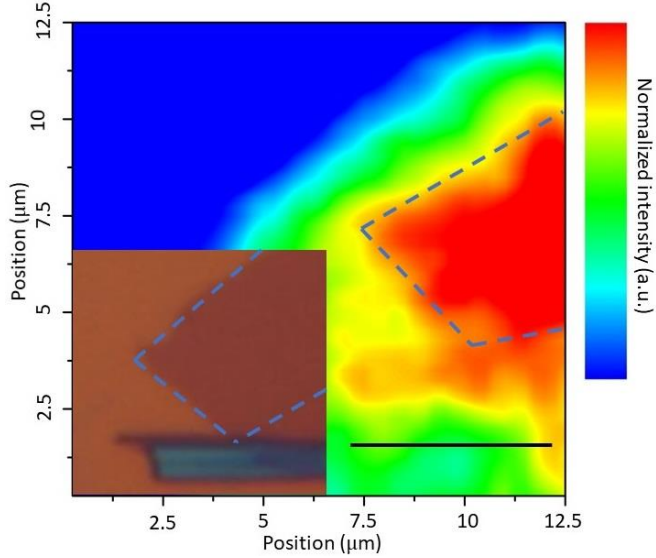

**Figure S8 | Cryogenic PL intensity mapping of trilayer heterostructure with 850 nm long pass filter.** The trilayer region is excited by 660 nm with 200 μW power (on sample) at 77 K. Blue dash line: trilayer sample region. Scale bar: ~5 μm.

### S9. Mass-action model analysis of interlayer exciton trion concentration

From Figure S4, the linewidth of interlayer exciton and trion PL are almost constant with temperature change. So the integrated PL of the two peaks using Lorentz fitting is proportional to the peak PL intensity. We can therefore relate the PL peak intensity ratio (Figure 3e) to the ratio of concentrations of trions to excitons and use the mass action model to observe:

$$\frac{n_{X-}}{n_X} \sim \frac{n_e \exp(\frac{E_T}{k_B T})}{k_B T} \tag{S3}$$

where $n_{X-}$, $n_X$ and $n_e$ are the concentration of natural excitons, trions and free electrons. $E_T$ is the trion binding energy. The variations of the three variables with temperature and the doping level are reported by Ross *et al* [S14]. $n_B = n_{X-} + n_e$ is the doping level and should be a constant. At non-zero doping levels then $n_{X-}$ and $n_e$ will show a sharp increase at low temperature and $n_X$ will be small. The $\frac{n_{X-}}{n_X}$ ratio would therefore show a sudden increase at low temperature as shown in Figure 3e. The abrupt rise around 10 K is a key feature of trions as mentioned in the main text. And that also agrees with the abrupt change in $n_{X-}$ and $n_X$ as predicted by Ross *et al* [S14]. Notice that the abrupt change only occurs when there is sufficient doping level so that there is large formation of trions.

### Supplementary References